\documentclass{article}
\usepackage{graphicx}
\usepackage{amsmath}
\usepackage{amssymb}
\usepackage{color}

\begin{document}

\vspace*{3cm} \thispagestyle{empty}
\vspace{5mm}

\noindent \textbf{\Large General Transformation Formulas for Fermi-Walker Coordinates}\\

\noindent \textbf{\normalsize David Klein}\footnote{Department of Mathematics, California State 
University, Northridge, Northridge, CA 91330-8313. Email: david.klein@csun.edu.} \textbf{\normalsize and Peter Collas}\footnote{Department of Physics and Astronomy, 
California State University, Northridge, Northridge, CA 91330-8268. Email: peter.collas@csun.edu.}
\\

\vspace{4mm} \parbox{11cm}{\noindent{\small We calculate the transformation and inverse transformation, in the form of Taylor expansions, from arbitrary coordinates to Fermi-Walker coordinates in tubular neighborhoods of arbitrary timelike paths for general spacetimes. Explicit formulas for coefficients and the Jacobian matrix are given.}\vspace{5mm}
}\\
\vspace{6cm}
\pagebreak

\setlength{\textwidth}{27pc}
\setlength{\textheight}{43pc}
\noindent \textbf{{\normalsize 1. Introduction}}\\

\noindent Measurements in a gravitational field are most easily interpreted through the use of a system of locally inertial coordinates. For an observer following a timelike worldline, Fermi-Walker coordinates provide such a system. A Fermi-Walker coordinate frame is nonrotating in the sense of Newtonian mechanics and is realized physically as a system of gyroscopes \cite{walker, synge}.  The metric along the path is Minkowskian, with first order corrections away from the path that depend only on the acceleration of the observer \cite{MTW}.  If the worldline is geodesic, the coordinates are commonly referred to as Fermi or Fermi normal coordinates, and the metric is Minkowskian to first order near the path with second order corrections due to curvature \cite{MM63}. Applications of these coordinate systems are voluminous. They include the study of tidal dynamics, gravitational waves, statistical mechanics, and quantum gravity effects  \cite{Chic, Ishii, Marz, FG, CK, B, PP82}. \\

\noindent Under general conditions, a timelike path has a neighborhood on which a Fermi-Walker coordinate system can be defined \cite{oniell}.  Some general results are known for expansions of the metric in these coordinates. In \cite{LN79+} Li and Ni derived the third order expansion of the metric and second order expansion of the equations of motion in Fermi-Walker coordinates for general spacetimes, and in \cite{LN79} they found expansions of the connection coefficients, metric, and geodesic equations in Fermi coordinates to third order, fourth order, and third order respectively, and gave an iteration scheme for calculation to higher order.  Marzlin investigated weak gravitational fields in \cite{Marz} and found the  expansion of the Minkowski metric with small perturbations to infinite order in Fermi-Walker coordinates.\\

\noindent For particular spacetimes and special timelike paths, some explicit transformation formulas are known.  For example, in \cite{Chic}, exact coordinate transformations were constructed for specific paths in de Sitter and G\"odel spacetimes, but the calculations for those examples were possible only because exact solutions for certain spacelike geodesics could be obtained in closed form.  This is not possible in general. In \cite{bini05} the coordinate transformation mapping the Kerr metric written in standard Boyer-Lindquist coordinates to its corresponding form in Fermi-Walker coordinates was approximated for a path with fixed space coordinates and then generalized to circular paths, but the coordinate transformation in the reverse direction was not provided.\\

\noindent To our knowledge, completely general and easily usable transformation formulas to and from Fermi-Walker (and therefore Fermi) coordinates do not appear in the literature.  This paper fills that gap.  We calculate an explicit formula for the general transformation, and inverse transformation, from \textit{a priori} coordinates to Fermi-Walker coordinates for arbitrary spacetimes, in the form of Taylor expansions.  The expansions are valid in sufficiently small neighborhoods of any timelike path.  In one direction, from \textit{a priori} coordinates to Fermi-Walker coordinates, the coefficients for the $n+2$ order terms involve $n$-th order partial derivatives of connection coefficients along the given timelike path in Fermi-Walker coordinates.  Thus, using the results of \cite{LN79+} and \cite{LN79}, the transformation law we provide in this direction is immediately available up fourth order for Fermi-Walker coordinates and to fifth order for Fermi coordinates (in the case that the timelike path is geodesic).  The transformation formula for the other direction, given by Theorem 3 below, from Fermi-Walker to a given coordinate system, is completely self-contained and is exact in the case that the coordinate transformation from Fermi-Walker coordinates is real analytic.  We elaborate on this matter in Section 4.  Our methods are more direct than those of \cite{bini05}, which require the solutions of systems of a large number of equations dependent on metric coefficients. Our method may in principle be used to calculate coordinate expansions and Jacobians to arbitrarily high order and is completely general.\\

\noindent In Section 2 we introduce notation and define Fermi-Walker coordinates.  In Section 3, using the fact that covariant derivatives of coordinate 1-forms are tensors, we calculate the Jacobian matrix for the transformation from Fermi-Walker coordinates to \textit{a priori} coodinates.  Section 4 gives the general transformations laws in both directions.  In Section 5 we apply the general method to develop coordinate transformation formulas for vector fields, and Killing fields, in particular. Section 6 illustrates the use of our formulas with examples.  Section 7 gives concluding remarks.\\

\noindent \textbf{{\normalsize 2. Fermi-Walker coordinates}}\\

\noindent Let $(M, g)$ be a four-dimensional Lorentzian $C^{n}$ manifold. For convenience we assume that $n\geq 4$,  but if necessary the results that follow may be readily adapted for smaller $n$.  The Levi-Civita connection is denoted by $\nabla$, and throughout we use the sign conventions of Misner, Thorne and Wheeler \cite{MTW}. A timelike path is a smooth map from an open interval on the real line to $M$, whose tangent vector is timelike.  A vector field $X$ is said to be Fermi-Walker transported along a timelike path $\sigma$ if $X$ satisfies the Fermi-Walker equations, which in coordinate form are given by,

\begin{equation}
F_{\vec{u}}X^{\alpha}\equiv \nabla_{\vec{u}}\;X^{\alpha}+\Omega^{{\alpha}}_{\;\,\beta}X^{\beta}=0\,\label{F1}.
\end{equation}
 
\noindent Here  $\vec{u}$ is the four-velocity along $\sigma$ (i.e., the unit tangent vector), $\Omega^{{\alpha}}_{\;\,\beta}=a^{\alpha}u_{\beta}- 
u^{\alpha} a_{\beta}$, and $a^{\alpha}$ is the four-acceleration.  As usual greek indices run over $0,1,2,3$ and lower case latin over
$1,2,3$. It is well-known and easily verified that $F_{\vec{u}}(\vec{u})=\vec{0}$, and if vector fields $X$ and $Y$ are Fermi-Walker transported along $\sigma$, the scalar product $X^{\beta}Y_{\beta}$ is invariant along $\sigma$.  Thus, a tetrad of vectors, Fermi-Walker transported along  $\sigma$ and orthonormal at one point on $\sigma$, is necessarily orthonormal at all points on the path. Moreover, such tetrads may be constructed so that one of the orthonormal vectors is the tangent vector $\vec{u}$. \\

\noindent Let $\sigma (\tau)$ denote the parameterization of $\sigma$ by proper time $\tau$, 
and let $e_{0}(\tau)$, $e_{1}(\tau), e_{2}(\tau), e_{3}(\tau)$ be an orthonormal Fermi-Walker transported tetrad along $\sigma$, with $e_{0}=\vec{u}$. We note that a method for constructing Fermi-Walker tetrads along a timelike path from their initial values is given in \cite{maluf}. The Fermi-Walker coordinates $x^{0}$, $x^{1}$, $x^{2}$, $x^{3}$ relative to the tetrad on $\sigma$ are given by,

\begin{equation}\label{F2}
\begin{split}
x^{0}\left (\exp_{\sigma(\tau)} (\lambda^{j}e_{j}(\tau)\right)&= \tau \\
x^{K}\left (\exp_{\sigma(\tau)} (\lambda^{j}e_{j}(\tau)\right)&= \lambda^{K}, 
\end{split} 
\end{equation}

\noindent where exponential map, $\exp_{p}(\vec{v})$, denotes the evaluation at affine parameter $1$ of the geodesic starting at the point $p$ in the spacetime, with initial derivative $\vec{v}$, and it is assumed that the $\lambda^{j}$ are sufficiently small so that the exponential maps in \eqref{F2} are defined. From the theory of differential equations, a solution to the geodesic equations depends smoothly on its initial data so it follows from Eq. \eqref{F2} that Fermi-Walker coordinates are smooth. Moreover, it follows from \cite{oniell} that there exists a neighborhood $U$ of $\sigma$ on which the map $ (x^{0}, x^{1}, x^{2}, x^{3}): U \rightarrow \mathbb{R}^{4}$ is well-defined, and it is a diffeomorphism onto the image of $U$. We refer to such a map as a Fermi-Walker coordinate chart $(x^{A}, U)$ for $\sigma$. By construction, it is a nonrotating coordinate system for the observer $\sigma$ \cite{walker, MTW, CK2}.\\

\noindent Let $\{y^{\alpha}\}$ be an arbitrary coordinate system on $M$ defined on an open set containing a portion (or all) of the timelike path $\sigma$. We refer to $\{y^{\alpha}\}$ as \textit{a priori} coordinates. We assume that the metric tensor is known in the \textit{a priori} coordinates, and that connection coefficients may therefore be readily computed in these coordinates. Henceforth, we use Greek indices and lower case Latin indices exclusively for the \textit{a priori} coordinates. In addition, we adopt the convention that the indices $A,B,C,D, E$ take the values $0,1,2,3$, while the indices $I,J,K,L$ are restricted to $1,2,3$, and we use these upper case Latin indices exclusively for Fermi-Walker coordinates. Following this notation, the Fermi-Walker tetrad vectors,  $e_{0}(\tau), e_{1}(\tau), e_{2}(\tau), e_{3}(\tau)$, along $\sigma$ are given by, 

\begin{equation}
e^{\alpha}_{A} = e^{\alpha}_{A}(\tau) = \frac{\partial y^{\alpha}}{\partial x^{A}}\Bigr|_{\sigma}\,,
\end{equation}

\noindent where the right side is evaluated at $x^{0}= \tau$ and $x^{K}= 0$.  The four-by-four matrix $e^{\alpha}_{A}(\tau)$ is thus the restriction to $\sigma$ of the Jacobian of the coordinate transformation $y^{\alpha}=y^{\alpha}(x^{A})$. The inverse of this Jacobian matrix is given by

\begin{equation}
e^{A}_{\alpha} = e^{A}_{\alpha}(\tau) \equiv \frac{\partial x^{A}}{\partial  y^{\alpha}}\Bigr|_{\sigma}\label{e4}.
\end{equation}

\noindent Finally, we mention that the non zero connection coefficients in Fermi-Walker coordinates, evaluated on $\sigma$, are given by \cite{MTW}:

\begin{equation}
\Gamma^{0}_{\;\,K0}=\Gamma^{K}_{\;\,00} = a^{K} \label{F4}.
\end{equation}

\noindent In the case that $\sigma$ is a geodesic so that $\vec{a} = 0$, all connection coefficients on $\sigma$ vanish.  However, partial derivatives of connection coefficients, with respect to Fermi-Walker coordinates, on $\sigma$ are in general not zero. In the case of Fermi coordinates, these derivatives along with expansion of the metric tensor to second order in the space variables, were computed in \cite{MM63,P80}.  Higher order derivatives and higher order expansions of the metric in Fermi and Fermi-Walker coordinates are given in \cite{LN79+, LN79, Ishii}.\\ 

\noindent \textbf{{\normalsize 3. The Jacobian}}\\

\noindent In this section, we begin by calculating Taylor polynomials, centered at a particular point on the timelike path $\sigma(\tau)$, for 1-form fields, in the Fermi-Walker variables, $x^{0}$, $x^{1}$, $x^{2}$, $x^{3}$.  Without loss of generality, we may expand about the point $\sigma(\tau=0)$.  The Taylor expansion for a 1-form field $V_{A}(x^{0},x^{1},x^{2},x^{3})$ has the form,

\begin{equation}
V_{A}(x^{0},x^{1},x^{2},x^{3})=V_{A}+x^{B}\frac{\partial V_{A}}{\partial x^{B}}+\frac{1}{2}x^{C}x^{B}\frac{\partial^{2}V_{A}}{\partial x^{C}\partial x^{B}}+ \cdots \label{y1}
\end{equation}

\noindent where, on the right side, $V_{A}$ and its derivatives are evaluated at $(\tau,0,0,0)$, with $\tau=0$.  Here and in what follows the ellipsis indicates either an infinite sum or a finite sum with remainder (in the case that the field $V$ is smooth but not analytic).\\

\noindent The $0th$ order terms in Eq. \eqref{y1} may be calculated directly from the Jacobian on $\sigma$,

\begin{equation}
V_{A}(\tau, 0,0,0)= e_{A}^{\alpha}(\tau)\,V_{\alpha}(\sigma (\tau))\label{x2'}
\end{equation}

\noindent Formulas for the higher order terms in \eqref{y1} may be deduced from the fact that covariant derivatives and multiple covariant derivatives of tensors are tensors,

\begin{eqnarray}
\nabla_{B}V_{A}\Bigr|_{\sigma} &=& e_{B}^{\beta}\,e^{\alpha}_{A}\,\nabla_{\beta}V_{\alpha}\label{x3}\\
\nabla_{C}\nabla_{B}V_{A}\Bigr|_{\sigma} &=& e_{C}^{\gamma}\,e_{B}^{\beta}\,e_{A}^{\alpha}\,\nabla_{\gamma}\nabla_{\beta}V_{\alpha}\label{x4}\\ &\cdots&\nonumber
\end{eqnarray}

\noindent with analogous third and higher covariant derivative expressions.  Thus, from Eq. \eqref{x3} it follows immediately that,

\begin{equation}
\frac{\partial V_{A}}{\partial x^{B}}(\tau, 0,0,0)=e_{B}^{\beta}\,e_{A}^{\alpha}\,\nabla_{\beta}V_{\alpha}+ \Gamma^{C}_{AB} e_{C}^{\gamma}V_{\gamma}\label{x5'},
\end{equation}

\noindent Similarly, from Eq. \eqref{x4},

\begin{equation}\label{x6}
\begin{split}
\frac{\partial^{2}V_{A}}{\partial x^{C}\partial x^{B}}(\tau,0,0,0)&= e_{C}^{\gamma}\,e_{B}^{\beta}\,e_{A}^{\alpha}\,\nabla_{\gamma}\nabla_{\beta}V_{\alpha}+\Gamma^{D}_{AB,C}V_{D}+\Gamma^{D}_{AB}\frac{\partial V_{D}}{\partial x^{C}}\\
&\quad +\Gamma^{D}_{AC}\nabla_{D}V_{B}+\Gamma^{D}_{CB}\nabla_{A}V_{D}
\end{split}
\end{equation}

\noindent Combining Eq. \eqref{x6} with Eq. \eqref{x5'} gives, 

\begin{equation}\label{x6'}
\begin{split}
\frac{\partial^{2}V_{A}}{\partial x^{C}\partial x^{B}}(\tau,0,0,0)&= e_{C}^{\gamma}\,e_{B}^{\beta}\,e_{A}^{\alpha}\,\nabla_{\gamma}\nabla_{\beta}V_{\alpha}+\Gamma^{D}_{AB,C}\, e_{D}^{\delta}V_{\delta}\\
&\quad +\Gamma^{D}_{AB}[e_{C}^{\gamma}\,e_{D}^{\delta}\,\nabla_{\gamma}V_{\delta}+ \Gamma^{E}_{DC} \,e_{E}^{\epsilon}V_{\epsilon}]\\
&\quad +\Gamma^{D}_{AC}\,e_{D}^{\delta}\,e_{B}^{\beta}\nabla_{\delta}V_{\beta}+\Gamma^{D}_{CB}\,e_{A}^{\alpha}\,e_{D}^{\delta}\nabla_{\alpha}V_{\delta}
\end{split}
\end{equation}

\noindent where $\Gamma^{A}_{BD,C}\equiv \frac{\partial}{\partial x^{C}} \Gamma^{A}_{BD}$ and all terms on the right side are evaluated at $\sigma (\tau)$. Explicit formulas for $nth$ order partial derivatives with respect to Fermi-Walker coordinates of the 1-form field $\{V_{A}\}$ may be similarly obtained in terms of $nth$ order covariant derivatives in the \textit{a priori} coordinate system $\{y^{\alpha}\}$ and $n-1st$ and lower order derivatives of the connection coefficients with respect to Fermi-Walker coordinates at the point $\sigma(\tau)$.  Thus, the Taylor coefficients in Eq. \eqref{y1} are given by formulas in the a priori coordinates on $\sigma$.\\  

\noindent \textbf{Theorem 1}. In a neighborhood of a point $\sigma(\tau)$ on the timelike path $\sigma$, the Jacobian of the transformation from Fermi-Walker coordinates $(x^{0},x^{1},x^{2},x^{3})$ to \textit{a priori} coordinates $(y^{0},y^{1},y^{2},y^{3})$ is given by,

\begin{equation} \label{j2}
\begin{split}
J^{\alpha}_{A}(x)&\equiv \frac{\partial y^{\alpha}}{\partial x^{A}}(x^{0},x^{1},x^{2},x^{3})\\&= e_{A}^{\alpha} + x^{B}\biggl\{ \Gamma^{C}_{AB} e_{C}^{\alpha}-e_{B}^{\beta}\,e_{A}^{\eta}\,\Gamma_{\beta \eta}^{\alpha} \biggr\}\\ 
&\quad + \frac{1}{2}x^{C}x^{B}
\biggl\{e_{C}^{\gamma}\,e_{B}^{\beta}\,e_{A}^{\eta}\left(\Gamma_{\gamma \eta}^{\mu}\Gamma_{\beta \mu}^{\alpha}+\Gamma_{\gamma \beta}^{\mu}\Gamma_{\mu\eta}^{\alpha}-\Gamma_{\beta \eta, \gamma}^{\alpha}\right)\\
&\quad +e_{D}^{\alpha}\Gamma^{D}_{AB,C}  + \Gamma^{D}_{AB}\left(\Gamma^{E}_{DC} \,e_{E}^{\alpha}-e_{C}^{\gamma}\,e_{D}^{\eta}\,\Gamma_{\gamma\eta}^{\alpha}\right)\\
&\quad -\Gamma^{D}_{AC}\,e_{D}^{\delta}\,e_{B}^{\beta}\Gamma_{\beta\delta}^{\alpha}-\Gamma^{D}_{CB}\,e_{A}^{\mu}\,e_{D}^{\eta}\Gamma_{\mu\eta}^{\alpha} \biggr \}+\cdots
\end{split}
\end{equation}

\noindent \textbf{Proof.} Without loss of generality, take $\tau=0$. The Taylor expansion given by Eq. \eqref{y1} is valid in some neighborhood $B_{\sigma(0)}$ of $\sigma(0)$.  Let a point $p \in B_{\sigma(0)}$ have  \textit{a priori} coordinates $(y^{0},y^{1},y^{2},y^{3})$ and let $V$ be a 1-form field on $B_{\sigma(0)}$ whose components relative to $\{y^{\alpha}\}$ are $V_{\alpha}(y^{0},y^{1},y^{2},y^{3})$. Corresponding to the \textit{a priori} coordinates $(y^{0},y^{1},y^{2},y^{3})$ there is a unique set of Fermi-Walker coordinates $(x^{0},x^{1},x^{2},x^{3})$.  Then by virtue of this correspondence, Eq. \eqref{y1} determines a map $J$ which transforms the components $V_{\alpha}(y^{0},y^{1},y^{2},y^{3})$ of $V$ at  $p$ to the Fermi-Walker components $V_{A}(x^{0},x^{1},x^{2},x^{3})$.\\

\noindent We may in particular apply the map $J$ to each of the following elements of the canonical basis of the contangent space at $p$:

\begin{eqnarray} \label{y2}
\begin{split}
dy^{0}\Bigr|_{p}&=(1,0,0,0)\\
dy^{1}\Bigr|_{p}&=(0,1,0,0),\\
dy^{2}\Bigr|_{p}&=(0,0,1,0),\\
dy^{3}\Bigr|_{p}&=(0,0,0,1)
\end{split}
\end{eqnarray}

\noindent Eq. \eqref{y1} may be used to compute $J(dy^{\alpha}\Bigr|_{p})$ to find the Fermi-Walker coordinates of $dy^{\alpha}\Bigr|_{p}$, that is, the $\alpha$-th row, $(\partial y^{\alpha}/\partial x^{0}, \partial y^{\alpha}/\partial x^{1}, \partial y^{\alpha}/\partial x^{2}, \partial y^{\alpha}/\partial x^{3}) $, of the Jacobian matrix of the transformation $y^{\alpha}= y^{\alpha}(x^{0},x^{1},x^{2},x^{3})$.  In matrix form, the $A$-th component, $J^{\alpha}_{A}(x)$, at $(x^{0},x^{1},x^{2},x^{3})$ of $J(dy^{\alpha}\Bigr|_{p})$ is given by,

\begin{equation}
J^{\alpha}_{A}(x) = \frac{\partial y^{\alpha}}{\partial x^{A}}(x^{0},x^{1},x^{2},x^{3})\label{j1}
\end{equation}

\noindent Now, setting $V_{\eta}= \delta^{\alpha}_{\eta}$ (the delta function) in Eqs. \eqref{x2'}, \eqref{x5'}, \eqref{x6'}, and \eqref{y1}, so that $\nabla_{\beta}V_{\eta} = -\Gamma_{\beta \eta}^{\alpha}(\sigma(0))$, yields Eq. \eqref{j2} for the Jacobian matrix.$\blacksquare$\\

\noindent Using the results of the next section, the Jacobian matrix given \eqref{j2}, and its inverse, may be used to transform vector fields  Fermi-Walker coordinates\\

\noindent \textbf{{\normalsize 4. Transformation of coordinates}}\\

\noindent In this section, we use the Jacobian \eqref{j2} to find coordinate transformations of the form $y^{\alpha}= y^{\alpha}(x^{0},x^{1},x^{2},x^{3})$ and $x^{A}= x^{A}(y^{0},y^{1},y^{2},y^{3})$.  Let the Taylor expansion for $y^{\alpha}$ be given by,

\begin{equation}
y^{\alpha}(x^{0},x^{1},x^{2},x^{3}) = y^{\alpha}_{0}+ b_{A}^{\alpha}x^{A}+c_{AB}^{\alpha}x^{A}x^{B}+d_{ABC}^{\alpha}x^{A}x^{B}x^{C}+\cdots \label{t1}
\end{equation}

\noindent where $y^{\alpha}_{0}= y^{\alpha}(0,0,0,0) = y^{\alpha}(\sigma(0))$. Taking partial derivatives of both sides of Eq. \eqref{t1} with respect to $x^{A}$ and comparing with Eq. \eqref{j2} yields the following coefficients to third order,

\begin{eqnarray}\label{t2}
\begin{split}
b_{A}^{\alpha}&= e_{A}^{\alpha}\\
2c_{AB}^{\alpha}&= \Gamma^{C}_{AB} e_{C}^{\alpha}-e_{B}^{\beta}\,e_{A}^{\eta}\,\Gamma_{\beta \eta}^{\alpha}\\
3!d_{ABC}^{\alpha}&= e_{C}^{\gamma}\,e_{B}^{\beta}\,e_{A}^{\eta}\left(\Gamma_{\gamma \eta}^{\mu}\Gamma_{\beta \mu}^{\alpha}+\Gamma_{\gamma \beta}^{\mu}\Gamma_{\mu\eta}^{\alpha}-\Gamma_{\beta \eta, \gamma}^{\alpha}\right)\\
&\quad +e_{D}^{\alpha}\Gamma^{D}_{AB,C}  + \Gamma^{D}_{AB}\left(\Gamma^{E}_{DC} \,e_{E}^{\alpha}-e_{C}^{\gamma}\,e_{D}^{\eta}\,\Gamma_{\gamma\eta}^{\alpha}\right)\\
&\quad -\Gamma^{D}_{AC}\,e_{D}^{\delta}\,e_{B}^{\beta}\Gamma_{\beta\delta}^{\alpha}-\Gamma^{D}_{CB}\,e_{A}^{\mu}\,e_{D}^{\eta}\Gamma_{\mu\eta}^{\alpha}
\end{split}
\end{eqnarray}

\noindent \textbf{Remark 1}.  Eq \eqref{t1} applied to $y^{\alpha}(x^{0},0,0,0)$ gives the expansion $\sigma(\tau) = \sigma(0) + \sigma' (0) \tau + \frac{1}{2}\sigma'' (0)\tau^{2}+\cdots$.  Similarly, the expansion for $y^{\alpha}(0,x^{1}s,0,0)$ gives coordinates for points lying on a spacelike geodesic orthogonal to $\sigma(\tau)$ at $\tau = 0$.\\

\noindent For the purpose of inverting the series \eqref{t1}, we employ of the following notation,

\begin{eqnarray}
X^{\alpha} &\equiv& e_{A}^{\alpha}x^{A} = b_{A}^{\alpha}x^{A}\label{n1}\\
Y^{\alpha} &\equiv& y^{\alpha}(x^{0},x^{1},x^{2},x^{3}) - y^{\alpha}_{0}\label{n2}\\
2C_{\beta \gamma}^{\alpha}&\equiv&2c_{AB}^{\alpha}e_{\beta}^{A}e_{\gamma}^{B}=\Gamma^{C}_{AB}\,e^{A}_{\beta}\,e^{B}_{\gamma}\,e_{C}^{\alpha}-\Gamma_{\beta \gamma}^{\alpha}\label{n3}\\
3!D_{\delta \lambda \eta}^{\alpha}&\equiv&3!d_{ABC}^{\alpha}e_{\delta}^{A}e_{\lambda}^{B}e_{\eta}^{C}\nonumber\\
&=&-\Gamma^{\alpha}_{\delta\lambda ,\eta}+\Gamma^{\alpha}_{\mu\lambda}\Gamma^{\mu}_{\delta\eta}+\Gamma^{\alpha}_{\delta\mu}\Gamma^{\mu}_{\lambda\eta}+\Gamma^{D}_{AB,C}\, e_{\delta}^{A}\,e_{\lambda}^{B}\,e_{\eta}^{C}\,e_{D}^{\alpha}\nonumber\\
&\quad+&\Gamma^{D}_{AB}\left[-e^{A}_{\delta}\,e^{B}_{\lambda}\,e_{D}^{\mu}\Gamma^{\alpha}_{\mu\eta}+ \Gamma^{E}_{DC} \,e_{\delta}^{A}\,e_{\lambda}^{B}\,e_{\eta}^{C}\,e_{E}^{\alpha}\right]\nonumber\\
&\quad-&\left[\Gamma^{D}_{AC}\Gamma^{\alpha}_{\lambda\mu}\,e^{A}_{\delta}\,e^{C}_{\eta}+\Gamma^{D}_{CB}\Gamma^{\alpha}_{\mu\delta}\,e^{B}_{\lambda}\,e^{C}_{\eta}\right]\,e_{D}^{\mu}\label{n4}
\end{eqnarray}

\noindent with analogous definitions for higher order coefficients.  Note that the coefficients $C_{\beta \gamma}^{\alpha}$, $D_{\delta \lambda \eta}^{\alpha}$, etc. are symmetric in the subscript indices. Eq. \eqref{t1} can now be rewritten in the form of a series that is easily invertible,

\begin{equation}
Y^{\alpha} = X^{\alpha}+ C_{\beta \gamma}^{\alpha}X^{\beta}X^{\gamma}+D_{\delta \lambda \eta}^{\alpha}X^{\delta}X^{\lambda}X^{\eta}+\cdots \label{t3}
\end{equation}

\noindent It is readily verified that the Taylor expansion for $X^{\alpha}$ is,

\begin{equation}
X^{\alpha} = Y^{\alpha} - C_{\beta \gamma}^{\alpha}Y^{\beta}Y^{\gamma}+ (2C_{\beta \lambda}^{\alpha}C_{\delta \eta}^{\beta}-D_{\delta \lambda \eta}^{\alpha})Y^{\delta}Y^{\lambda}Y^{\eta} +\cdots \label{t5}
\end{equation}

\noindent Thus, from Eqs. \eqref{n1} and \eqref{n2} we may write,

\begin{equation}\label{t6}
\begin{split}
x^{A} &= e^{A}_{\alpha}(y^{\alpha}- y^{\alpha}_{0})- e^{A}_{\alpha}C_{\beta \gamma}^{\alpha}(y^{\beta}- y^{\beta}_{0})(y^{\gamma}- y^{\gamma}_{0})\\
&\quad+ e^{A}_{\alpha}(2C_{\beta \lambda}^{\alpha}C_{\delta \eta}^{\beta}-D_{\delta \lambda \eta}^{\alpha})(y^{\delta}- y^{\delta}_{0})(y^{\lambda}- y^{\lambda}_{0})(y^{\eta}- y^{\eta}_{0}) +\cdots
\end{split} 
\end{equation}

\noindent \textbf{Remark 2}. Eq. \eqref{t6} may be easily recast as an expansion about any fixed point $\sigma(\tau_{0})$ on $\sigma(\tau)$.  This is accomplished by redefining $e^{A}_{\alpha} \equiv e^{A}_{\alpha}(\tau_{0})$ and $y^{\alpha}_{0} \equiv y^{\alpha} (\sigma(\tau_{0}))$ in Eqs. \eqref{t2}, \eqref{n1} - \eqref{n4}, \eqref{t6}, and replacing $x^{0}$ by  $x^{0}-\tau_{0}$ in Eq. \eqref{t6}.\\

\noindent For Theorem 2 below, we now make the assumption that the tangent vector $\partial /\partial y^{0}$ is timelike in $U$ so that $y^{0} \equiv t$ may be selected as a time coordinate in the the \textit{a priori} coordinate system $(t,y^{1},y^{2},y^{3})$. We assume further that $\tau$ is an increasing function $\tau(t)$ of $t$ along $\sigma$ so that $\sigma$ may be parameterized by $t$.  This excludes causality violations along $\sigma$. Employing a standard abuse of notation, we write $\sigma(t)$ for this parameterization of $\sigma$ (as opposed to $\sigma(\tau(t))$). Similarly, the Fermi-Walker tetrad $e_{\alpha}(\tau)$ may be reparameterized by $t$ and we denote that parameterization by $e_{\alpha}(t)$ and Eqs. \eqref{n3} and \eqref{n4} are correspondingly modified.\\

\noindent \textbf{Theorem 2}. With the notation and assumptions of the preceding paragraph, the Taylor expansion for the transformation from \textit{a priori} coordinates to Fermi-Walker coordinates in a neighborhood of $\sigma$ is given by,
\begin{equation}\label{t7}
\begin{split}
x^{A}(t,y^{1},y^{2}&, y^{3})= \tau(t) \delta_{0}^{A}\\
&+e^{A}_{k}(t)(y^{k}- y^{k}_{0}(t))- e^{A}_{\alpha}(t)C_{jk}^{\alpha}(y^{j}- y^{j}_{0}(t))(y^{k}- y^{k}_{0}(t))\\
&+ e^{A}_{\alpha}(t)(2C_{\beta j}^{\alpha}C_{i k}^{\beta}-D_{i j k}^{\alpha})(y^{i}- y^{i}_{0}(t))(y^{j}- y^{j}_{0}(t))(y^{k}- y^{k}_{0}(t))\\
& +\cdots,
\end{split} 
\end{equation}

\noindent where as before $i,j,k=1,2,3$, and $y^{\alpha}_{0}(t) \equiv y^{\alpha} (\sigma(t))$.\\

\noindent \textbf{Proof.} For a given point $p \in U$ with \textit{a priori} coordinates $(t,y^{1},y^{2},y^{3})$, Eq. \eqref{t6} may be revised so as to be an expansion about the point $\sigma(t)$.  This is accomplished by redefining $e^{A}_{\alpha} \equiv e^{A}_{\alpha}(t)$ and $y^{\alpha}_{0} \equiv y^{\alpha} (\sigma(t))$ in Eqs. \eqref{t2} - \eqref{n4}.  Since $y^{0}_{0} = t = y^{0}$, Eq. \eqref{t6} becomes a polynomial in $y^{1},y^{2},y^{3}$ with coefficients that depend on $t$ yielding Eq. \eqref{t7}.$\blacksquare$\\

\noindent \textbf{Remark 3}. For the purpose of numerical computations, the expressions for the coefficients in Eq. \eqref{t7}, as well as for Eqs. \eqref{j2}, \eqref{t1}, and \eqref{t6} may be simplified by using Eqs. \eqref{t10'}, \eqref{t11'}, and \eqref{gamma1} which appear below in the proof of Theorem 3.\\

\noindent \textbf{Remark 4}. Fermi-Walker coordinates along $\sigma(t)$ determine a foliation of a neighborhood of  $\sigma$ by space slices, each with constant $\tau = x^{0}$ coordinate. Given the coordinates $(t,y^{1},y^{2},y^{3})$ of a point $p$ near $\sigma(t)$, Eq. \eqref{t7} may be used to locate the space slice containing $p$ by estimating $x^{0}$.\\

\noindent We next find a series for the inverse transformation to Eq. \eqref{t7}. In Theorem 3 below, $\delta^{(\alpha)}_{\gamma}$ represents the coordinate 4-tuple for $dy^{\alpha}$ given by Eqs. \eqref{y2}. The parentheses enclosing the index $\alpha$ indicate that $\delta^{(\alpha)}_{\gamma}$ should be understood as a $(0,1)$ tensor in the calculations of the coefficients in \eqref{t13'} below, rather than as a $(1,1)$ tensor.\\

\noindent \textbf{Theorem 3}. For a point $p$ with Fermi-Walker coordinates $(\tau,x^{1},x^{2},x^{3})$ sufficiently close to the timelike path $\sigma(\tau)$, the \textit{a priori} coordinates $(y^{0},y^{1},y^{2},y^{3})$ of $p$ are given by,

\begin{equation}\label{t13'}
\begin{split}
y^{\alpha}&(\tau,x^{1},x^{2},x^{3})=  y^{\alpha}(\tau,0,0,0)+e_{J_{1}}^{\alpha}(\tau)x^{J_{1}}\\
&+\frac{1}{2}(\nabla_{\mu_{2}} \delta^{(\alpha)}_{\mu_{1}})e_{J_{1}}^{\mu_{1}}(\tau)e_{J_{2}}^{\mu_{2}}(\tau)x^{J_{1}}x^{J_{2}}\\
&+\frac{1}{3!}(\nabla_{\mu_{3}}\nabla_{\mu_{2}} \delta^{(\alpha)}_{\mu_{1}})e_{J_{1}}^{\mu_{1}}(\tau)e_{J_{2}}^{\mu_{2}}(\tau)e_{J_{3}}^{\mu_{3}}(\tau)x^{J_{1}}x^{J_{2}}x^{J_{3}} + \cdots \\
&+\frac{1}{n!}(\nabla_{\mu_{n}}\cdots\nabla_{\mu_{2}} \delta^{(\alpha)}_{\mu_{1}})e_{J_{1}}^{\mu_{1}}(\tau)\cdots e_{J_{n}}^{\mu_{n}}(\tau)x^{J_{1}}\cdots x^{J_{n}} + \cdots,
\end{split}
\end{equation}

\noindent where $y^{\alpha}(\tau,0,0,0)= \sigma^{\alpha}(\tau)$, and where in each term the sum on the $\mu_{k}$'s is to be carried out before the sum on the $J_{k}$'s. If $y^{\alpha}$ is an analytic function of $(\tau,x^{1},x^{2},x^{3})$, Eq. \eqref{t13'} is an infinite series. If not, it should be interpreted as a Taylor polynomial as indicated above.\\

\noindent Before proving Theorem 3 we give the following Corollary, which is an immediate consequence of Theorem 3.  However, we also include an independent, elementary proof to which we refer in the proof of Theorem 3.\\

\noindent \textbf{Corollary 1}. With the same assumptions as in Theorem 3, $y^{\alpha}$ may be expressed directly in terms of the \textit{a priori }connection coefficients as follows,

\begin{equation}\label{t8}
\begin{split}
y^{\alpha}(\tau,x^{1},x^{2},x^{3})= &\, y^{\alpha}(\tau,0,0,0)+e_{K}^{\alpha}(\tau)x^{K}\\
&-\frac{1}{2}\Gamma_{\beta\gamma}^{\alpha}(\sigma(\tau))e_{J}^{\beta}(\tau)e_{K}^{\gamma}(\tau)x^{J}x^{K}\\
&+\frac{1}{3!}\biggl\{2\Gamma_{\beta\mu}^{\alpha}(\sigma(\tau))\Gamma_{\gamma\delta}^{\mu}(\sigma(\tau))-\Gamma_{\beta\gamma,\delta}^{\alpha}(\sigma(\tau))\biggr\}\\
&\times e_{I}^{\delta}(\tau)e_{J}^{\beta}(\tau)e_{K}^{\gamma}(\tau)x^{I}x^{J}x^{K} + \cdots
\end{split}
\end{equation}

\noindent \textbf{Proof of Corollary 1}. Eqs. \eqref{F2} may be expressed as the  evaluation at $s=1$ of the solution $y^{\alpha}(s)$ of the initial value problem,

\begin{equation}
\begin{split}\label{t9}
\frac{d^{2}y^{\alpha}}{ds^{2}}+ \Gamma_{\beta\gamma}^{\alpha}\frac{dy^{\beta}}{ds}\frac{dy^{\gamma}}{ds}&=0\\
y^{\alpha}(0)&=\sigma^{\alpha}(\tau)\\
\frac{dy^{\alpha}}{ds}(0)&= x^{K}e_{K}^{\alpha}(\tau)
\end{split}
\end{equation}

\noindent From the initial conditions, the Taylor expansion for $y^{\alpha}(s)$ has the form,

\begin{equation}
y^{\alpha}(s)=\sigma^{\alpha}(\tau)+e_{K}^{\alpha}(\tau)x^{K}s+a_{2}^{\alpha}\,\frac{s^{2}}{2}+a_{3}^{\alpha}\,\frac{s^{3}}{3!}+\cdots\label{t10}
\end{equation}

\noindent Similarly, 

\begin{equation}
\Gamma_{\beta\gamma}^{\alpha}(y(s))=\Gamma_{\beta\gamma}^{\alpha}(\sigma(\tau))+\Gamma_{\beta\gamma, \delta}^{\alpha}(\sigma(\tau))(y^{\delta}(s)-\sigma^{\delta}(\tau))+\cdots\label{t11}
\end{equation}

\noindent In a standard way, substituting Eqs. \eqref{t10} and \eqref{t11} into the geodesic equation \eqref{t9}, to solve for coefficients yields, 

\begin{equation}
\begin{split}\label{t12}
y^{\alpha}(s;\tau,x^{1},x^{2},x^{3})=\quad &\sigma^{\alpha}(\tau)+s e_{K}^{\alpha}(\tau)x^{K}\\
&-\frac{s^{2}}{2}\Gamma^{\alpha}_{\beta\gamma}(\sigma(\tau))e_{J}^{\beta}(\tau)e_{K}^{\gamma}(\tau)x^{J}x^{K}\\
&+\frac{s^{3}}{3!}\biggl\{2\Gamma_{\beta\mu}^{\alpha}(\sigma(\tau))\Gamma_{\gamma\delta}^{\mu}(\sigma(\tau))-\Gamma_{\beta\gamma,\delta}^{\alpha}(\sigma(\tau))\biggr\}\\
&\times e_{I}^{\delta}(\tau)e_{J}^{\beta}(\tau)e_{K}^{\gamma}(\tau)x^{I}x^{J}x^{K}+\cdots
\end{split}
\end{equation}

\noindent The result now follows by setting $s=1$. $\blacksquare$\\

\noindent \textbf{Remark 5}. Using the methods of the preceding proof, it is straightforward to compute higher order terms in Eq. \eqref{t8}.  For a well-written treatment of the analogous development of Riemann normal coordinates see \cite{brewin1, brewin2}.\\ 

\noindent \textbf{Proof of Theorem 3}. The initial value problem \eqref{t9} may be reformulated in Fermi-Walker coordinates as,

\begin{equation}
\begin{split}\label{t9'}
\frac{d^{2}X^{A}}{ds^{2}}+ \Gamma_{BC}^{A}\frac{dX^{B}}{ds}\frac{dX^{C}}{ds}&=0\\
X(0)&=(\tau,0,0,0)\\
\frac{dX}{ds}(0)&=(0, x^{1}, x^{2} ,x^{3})
\end{split}
\end{equation}

\noindent where $\tau$ is fixed. The solution is a linear function of the affine parameter $s$ given by $X(s)= (\tau, sx^{1}, sx^{2}, sx^{3})$, which together with Eq. \eqref{t9'} yields,

\begin{equation}
 \Gamma_{IJ}^{A}x^{I}x^{J}=0\label{t10'}
\end{equation}

\noindent at any point on $\sigma$ and all choices of $(x^{1}, x^{2} ,x^{3})$. Since $\Gamma_{IJ}^{A}$ is symmetric in its two lower indices, and since Eq. \eqref{t10'} holds for all $x$, it follows that $\Gamma_{IJ}^{A}=0$, a fact already noted in the remarks preceding Eq. \eqref{F4}. However, differentiating the geodesic equation in \eqref{t9'} with respect to $s$ and using the linearity of the solution yields,

\begin{equation}
 \Gamma_{IJ,K}^{A}x^{I}x^{J}x^{K}=0\label{t11'}
\end{equation}

\noindent at any point on $\sigma$, and differentiating repeatedly yields the analogous higher order identities,

\begin{equation}
 \Gamma_{J_{1} J_{2},J_{3}\cdots J_{n}}^{A}x^{J_{1}}\cdots x^{J_{n}}=0\label{gamma1},
\end{equation}

\noindent where the comma indicates that the connection coefficient is differentiated with respect to $x^{J_{3}},\cdots ,x^{J_{n}}$, and the result is then evaluated at any point on $\sigma$.\\

\noindent For fixed $\tau$, the Taylor expansion for $y^{\alpha}(\tau,x^{1},x^{2},x^{3})$ is given by,

\begin{equation}
y^{\alpha}(\tau,x^{1},x^{2},x^{3})= y^{\alpha}(\tau,0,0,0)+\frac{\partial y^{\alpha}}{\partial x^{K}}(\tau,0,0,0)x^{K}+\cdots\label{taylor1}
\end{equation}

\noindent Let $\delta^{(\alpha)}_{\gamma}$ represent the coordinate 4-tuple for $dy^{\alpha}$ given by Eqs. \eqref{y2}, as described above.  Then $V^{(\alpha)}_{K}=J^{\gamma}_{K}\delta^{(\alpha)}_{\gamma}= \partial y^{\alpha}/\partial x^{K}$ gives the components of the same 1-form in Fermi-Walker coordinates.  Thus, Eq. \eqref{taylor1} may be written as,

\begin{equation}
\begin{split}\label{taylor2}
y^{\alpha}(\tau,x^{1},x^{2},x^{3})= &y^{\alpha}(\tau,0,0,0)+ V^{(\alpha)}_{K} (\tau,0,0,0)x^{K}\\
+&\frac{\partial V^{(\alpha)}_{K}}{\partial x^{J}}(\tau,0,0,0)x^{K}x^{J}\\+&\frac{1}{3!}\frac{\partial^{2}V^{(\alpha)}_{K}}{\partial x^{I}\partial x^{J}}(\tau,0,0,0)x^{I}x^{K}x^{J}+\cdots
\end{split}
\end{equation}

\noindent Combining Eqs. \eqref{x2'}, \eqref{x5'}, \eqref{x6} with Eqs. \eqref{t10'} and \eqref{t11'} gives the following formulas for the first, second, and third order terms in Eq. \eqref{taylor2},

\begin{equation}
\begin{split}\label{taylor3}
V^{(\alpha)}_{K} (\tau,0,0,0)x^{K}= &\,\delta^{(\alpha)}_{\mu_{1}}\,e_{K}^{\mu_{1}}(\tau)x^{K}=e_{K}^{\alpha}(\tau)x^{K}\\
\frac{\partial V^{(\alpha)}_{K}}{\partial x^{J}}(\tau,0,0,0)x^{J}x^{K} = &(\nabla_{\mu_{2}}\delta^{(\alpha)}_{\mu_{1}})\,e_{J}^{\mu_{2}}(\tau)e_{K}^{\mu_{1}}(\tau)x^{J}x^{K}\\
\frac{\partial^{2}V^{(\alpha)}_{K}}{\partial x^{I}\partial x^{J}}(\tau,0,0,0)x^{I}x^{K}x^{J}=&(\nabla_{\mu_{3}}\nabla_{\mu_{2}}\delta^{(\alpha)}_{\mu_{1}})e_{I}^{\mu_{3}}(\tau)e_{J}^{\mu_{2}}(\tau)e_{K}^{\mu_{1}}(\tau)x^{I}x^{J}x^{K}
\end{split}
\end{equation}

\noindent The analogous formula for the general $n$th order coefficients may be deduced using Eq. \eqref{gamma1} and the higher order analogs to Eq. \eqref{x4} as follows, 

\begin{equation}
\begin{split}\label{gamma2}
(\nabla_{J_{n}}\cdots \nabla_{J_{2}}V_{J_{1}})x^{J_{1}}&\cdots x^{J_{n}}= \left(\frac{\partial}{\partial x^{J_{n}}}\nabla_{J_{n-1}}\cdots \nabla_{J_{2}}V_{J_{1}}\right)x^{J_{1}}\cdots x^{J_{n}}\\
=&\left(\frac{\partial}{\partial x^{J_{n}}}\frac{\partial}{\partial x^{J_{n-1}}}\nabla_{J_{n-2}}\cdots \nabla_{J_{2}}V_{J_{1}}\right)x^{J_{1}}\cdots x^{J_{n}}\\
=&\qquad \cdots\\
=&\,\frac{\partial^{n-1}V_{J_{1}}}{\partial x^{J_{n}}\cdots \partial x^{J_{2}}}x^{J_{1}}\cdots x^{J_{n}},
\end{split}
\end{equation}

\noindent where all multiple derivatives and covariant derivatives are evaluated at $(\tau,0,0,0)$. $\blacksquare$\\

\noindent \textbf{Remark 6}. A derivation of the transformation formulas in this paper may be carried out in the reverse direction.  An alternative derivation of Eq. \eqref{t1}, and thereafter Eq. \eqref{j2}, is possible by starting with Eq. \eqref{t8} or \eqref{t13'}.  Eq. \eqref{t1} results by computing Taylor expansions in $\tau$ of the coefficients of Eq. \eqref{t8} and collecting terms.  For that purpose, Eq. \eqref{F1} applied to $e_{K}^{\alpha}(\tau)$ may be used to compute $de_{K}^{\alpha}(0)/d\tau$. These derivatives may be expressed in terms of connection coefficients in Fermi-Walker coordinates. \\

\noindent \textbf{{\normalsize 5. Vector Fields}}\\

\noindent The expansions of Sect. 3 for 1-form fields may be extended more generally to tensor fields.  If $T$ is a smooth tensor field defined in a neighborhood of the timelike path $\sigma$ in terms of the  \textit{a priori} coordinates $\{y^{\alpha}\}$, then we may write,
\begin{equation}\label{nx1}
\begin{split}
T(\tau, x^{1},x^{2},x^{3})&=T(\tau, 0,0,0)+x^{K}\frac{\partial T}{\partial x^{K}}(\tau, 0,0,0)\\ &\quad +\frac{1}{2}x^{J}x^{K}\frac{\partial^{2}T}{\partial x^{J}\partial x^{K}}(\tau, 0,0,0)+\cdots, 
\end{split}
\end{equation}

\noindent and the methods of Sect. 2 may be applied to compute the Taylor coefficients for the components of $T$. For ease of exposition, and to reveal a computational short cut,  we specialize to the case that $T$ is a vector field with components $V^{\beta}$ with respect to the coordinates $\{y^{\alpha}\}$.  The contravariant versions of Eqs. \eqref{x2'},\eqref{x3},\eqref{x4}, and the resulting analogs to Eqs. \eqref{x5'}, \eqref{x6}, and \eqref{x6'} yield the following result.\\

\noindent \textbf{Theorem 4}.  Let $\sigma = \sigma (\tau)$ be a timelike path parametrized by proper time $\tau = x^{0}$. Let $V = V^{\gamma}(y^{0},y^{1},y^{2},y^{3})$ be a vector field defined on an open set containing $\sigma$. Then in a sufficiently small neighborhood of $\sigma$, the components of $V$ in Fermi-Walker coordinates are given by,

\begin{eqnarray}
V^{A}(\tau,x^{1},x^{2},x^{3})
&=& e^{A}_{\gamma}V^{\gamma}+x^{K}\,\big[e^{A}_{\alpha}e_{K}^{\beta}\nabla_{\beta}V^{\alpha}-\Gamma_{KC}^{A}e^{C}_{\gamma}V^{\gamma}\big]\nonumber\\
&+& \frac{1}{2}x^{J}x^{K}\big[e_{J}^{\gamma}\,e_{K}^{\beta}\,e^{A}_{\alpha}\,\nabla_{\gamma}\nabla_{\beta}V^{\alpha}-\Gamma^{A}_{KD,J} e^{D}_{\delta}V^{\delta}\nonumber\\
&-&\Gamma^{A}_{KD}(e_{J}^{\beta}\,e_{\alpha}^{D}\,\nabla_{\beta}V^{\alpha}- \Gamma^{D}_{JC}e_{\gamma}^{C}V^{\gamma})\label{nM6}\\ 
&-&\Gamma^{A}_{JD}e^{D}_{\delta}e^{\gamma}_{K}\nabla_{\gamma}V^{\delta}+\Gamma^{D}_{JK} e^{A}_{\alpha}e^{\delta}_{D} \nabla_{\delta}V^{\alpha}\big]\nonumber\\
&+&O(3),\nonumber
\end{eqnarray}

\noindent where the coefficients on the right hand side are evaluated at $\sigma(\tau)$ and $O(3)$ indicates terms of order $3$ and higher in the space variables.\\ 

\noindent In the case of Fermi coordinates along a timelike geodesic, the connection coefficients vanish, and the result is a substantially simpler formula.\\

\noindent \textbf{Corollary 2}.  With the same assumptions as in Theorem 4 except that $\sigma =\sigma(\tau)$ is a timelike geodesic,
\begin{eqnarray}
V^{A}(\tau,x^{1},x^{2},x^{3})
&=&e^{A}_{\gamma}V^{\gamma}+x^{K}\,e^{A}_{\gamma}e_{K}^{\alpha}\nabla_{\alpha}V^{\gamma}\nonumber\\
&+& \frac{1}{2}x^{J}x^{K}\left(e_{J}^{\alpha}e_{K}^{\beta}e^{A}_{\gamma}\nabla_{\alpha}\nabla_{\beta}V^{\gamma}-V^{\gamma}e^{D}_{\gamma}\Gamma^{A}_{DK,J}\right)\label{nM6+}\\
&+&\ O(3).\nonumber
\end{eqnarray}

\noindent Killing fields play important roles in identifying a concept of energy in the context of general relativity, studying event horizons, and for other purposes.  If $V$ is a Killing vector field, then, $\nabla_{\alpha}\nabla_{\beta}V^{\gamma}=R^{\gamma}_{\;\beta\alpha\nu}V^{\nu}$.  Substituting this into Eq. \eqref{nM6+} yields the following.\\

\noindent \textbf{Corollary 3}.  With the same assumptions as in Theorem 4 except that $\sigma =\sigma(\tau)$ is a timelike geodesic and $V$ is a Killing vector field,
\begin{eqnarray}
V^{A}(\tau,x^{1},x^{2},x^{3})
&=&e^{A}_{\gamma}V^{\gamma}+x^{K}\,e^{A}_{\gamma}e_{K}^{\alpha}\nabla_{\alpha}V^{\gamma}\nonumber\\
&+& \frac{1}{2}x^{J}x^{K}\left(e_{J}^{\alpha}e_{K}^{\beta}e^{A}_{\gamma}R^{\gamma}_{\;\beta\alpha\nu}V^{\nu}-V^{\gamma}e^{D}_{\gamma}\Gamma^{A}_{DK,J}\right)\label{nM6++}\\
&+& O(3)\nonumber
\end{eqnarray}

\noindent Since $V^{\nu} = e^{\nu}_{D}e^{D}_{\mu}V^{\mu}$, the coefficients of the second order terms, $x^{J}x^{K}$, may be expressed in Fermi-Walker coordinates as 
$\frac{1}{2}(R^{A}_{\,\,\,KJD}V^{D}-\Gamma^{A}_{KD,J}V^{D})$.  From Eq. (21) of \cite{LN79}, we may replace $D$ by $L$ in this last expression, thus omitting the $D=0$ term, and then rewrite these coefficients of second order terms in Eq. \eqref{nM6++} as,
\begin{equation}
\begin{split}\label{nM7++}
 \frac{1}{2}\left(R^{A}_{\,\,\,KJL}-\frac{1}{2}\Gamma^{A}_{KL,J}\right)V^{L}x^{J}x^{K}&=\frac{1}{2}R^{A}_{\,\,\,KJL}x^{J}x^{K}V^{L}\\&+\frac{1}{6}\left(R^{A}_{\,\,\,KLJ}+R^{A}_{\,\,\,LKJ}\right)V^{L}x^{J}x^{K}\\
 &=\frac{1}{3}R^{A}_{\,\,\,KJL}x^{J}x^{K}V^{L},
\end{split}
\end{equation}

\noindent where in the last line we have used the curvature symmetry, $R^{A}_{\,\,\,LKJ}=R^{A}_{\,\,\,L[KJ]}$. The right side of Eq. \eqref{nM7++} may be used to replace the coefficient of the second order terms in Eq. \eqref{nM6++}. We note in particular that if the Killing vector $V$ is tangent to the geodesic $\sigma(\tau)$, then $V^{L}=0$ for $L=1,2,3$ and the second order terms in Eq. \eqref{nM6++} vanish, i.e., 
\begin{eqnarray}
V^{A}(\tau,x^{1},x^{2},x^{3})
=e^{A}_{\gamma}V^{\gamma}(\sigma(\tau))+x^{K}\,e^{A}_{\gamma}e_{K}^{\alpha}\nabla_{\alpha}V^{\gamma}(\sigma(\tau))+O(3).
\label{nM8++}
\end{eqnarray}

\noindent We include an illustration of this last result in the next section.\\

\noindent \textbf{{\normalsize 6. Examples}}\\

\noindent In this section we illustrate results of previous sections with four examples.  In Example 1, we apply Eqs. \eqref{t7} and \eqref{nM8++} to the case of a circular geodesic orbit, with a tangent Killing vector, around the central mass $M$ in Schwarzschild spacetime.  We use the results to compare surfaces of simultaneity in Schwarzschild and Fermi normal coordinates. In Example 2 we use Eq. \eqref{t8} to calculate $y^{\alpha}(\tau,x^{1},x^{2},x^{3})$ for a timelike trajectory with fixed space coordinates in the de Sitter universe, obtaining to $O(4)$ Chicone and Mashhoon's result in \cite{Chic}. In the third example, we compute coordinates relative to a Zero Angular Momentum Observer frame in Kerr spacetime. In Example 4, we compute a coordinate transformation to Fermi-Walker coordinates for an observer at fixed Boyer-Lindquist coordinates in Kerr spacetime.\\

\noindent \textbf{Example 1}.  We take as the \textit{a priori} coordinates the usual coordinates in Schwarzschild spacetime, those of the Schwarzschild observer, $y^{0}=t,\;y^{1}=r,\;y^{2}=\theta,\;y^{3}=\phi$, in which the metric is given as,
\begin{equation}
ds^{2}=-\left(1-\frac{2M}{r}\right)dt^{2}+\displaystyle\frac{dr^{2}}{\displaystyle
\left(1-\frac{2M}{r}\right)}+r^{2}(d\theta^{2}+\sin^{2}\theta d\phi^{2}).
\label{S1}
\end{equation}

\noindent In these coordinates, the circular geodesic orbit around the central mass, with radius corresponding to radial coordinate $r_{0}$, is given by,

\begin{equation}
\label{S2}
\sigma(t)=\left(t, r_{0}, \pi/2, \beta t\right),
\end{equation}

\noindent where $\beta \equiv \sqrt{M/r_{0}^{3}}$. We note that $r_{0}> 3M$ is necessary for a timelike geodesic orbit, and that stable orbits are possible only for $r_{0}> 6M$. Thus, in the language of Theorem 2, $y^{1}_{0}(t)=r_{0}$, $y^{2}_{0}(t)= \pi/2$, and $y^{3}_{0}(t)= \beta t$.\\  

\noindent It is easily computed that,

\begin{equation}
\label{S3}
\tau = \tau (t)=\sqrt{\frac{r_{0}-3M}{r_{0}}}t \equiv \alpha t.
\end{equation}

\noindent For ease of notation, let,
\begin{eqnarray}
X&=&1-\frac{2M}{r_{0}},\nonumber\\
\epsilon&=&\frac{r_{0}-2M}{\sqrt{r_{0}(r_{0}-3M)}},\label{S4}\\
l&=&r_{0}\sqrt{\frac{M}{r_{0}-3M}},\nonumber
\end{eqnarray}

\noindent so that $\epsilon$ is the energy per unit mass of a test particle on the orbit, and $l$ is its angular momentum per unit mass. A version of Eq. \eqref{e4} for this example is \cite{CK, PP82},

\begin{eqnarray}
e^{0}&=&\left (\epsilon,0,0,-l\right ),\label{S9}\\
e^{1}&=&\left (\frac{l\sqrt{X}\sin(\alpha\beta t)}{r_{0}},\frac{\cos(\alpha\beta t)}{\sqrt{X}},0,
-\frac{\epsilon r_{0}\sin(\alpha\beta t)}{\sqrt{X}}\right ),\label{S10}\\
e^{2}&=&\left (0,0,r_{0},0\right ),\label{S11}\\
e^{3}&=&\left (\frac{-l\sqrt{X}\cos(\alpha\beta t)}{r_{0}},\frac{\sin(\alpha\beta t)}{\sqrt{X}},0,
\frac{\epsilon r_{0}\cos(\alpha\beta t)}{\sqrt{X}}\right ),\label{S12}
\end{eqnarray}

\noindent where the ordering of components is given by $(t,r,\theta,\phi)$. Using Eqs. \eqref{n3} and \eqref{t7} we readily obtain the functions $x^{A}(t,r,\theta,\phi)$ to $O(2)$,
\begin{eqnarray}
x^{0}&=&\alpha t-\frac{l}{r_{0}}r\left(\phi -  \beta t \right) + \cdots,\label{S13}\\
x^{1}&=&e^{1}_{\,r}(r-r_{0})+\frac{e^{1}_{\,\phi}}{r_{0}}\left(\phi -  \beta t\right)\nonumber\\
&+&\frac{e^{1}_{\,r}}{2}\left[\Gamma^{r}_{\;rr}(r-r_{0})^{2}+\Gamma^{r}_{\;\theta\theta}(\theta-\pi/2)^{2}+\Gamma^{r}_{\;\phi\phi}(\phi-\beta t)^{2}\right]+ \cdots,\label{S14}\\
x^{2}&=&r(\theta-\pi/2)+ \cdots,\label{S15}\\
x^{3}&=&e^{3}_{\,r}(r-r_{0})+\frac{e^{3}_{\,\phi}}{r_{0}}r(\phi-\beta t)\nonumber\\
&+&
\frac{e^{3}_{\,r}}{2}\left[\Gamma^{r}_{\;rr}(r-r_{0})^{2}+\Gamma^{r}_{\;\theta\theta}(\theta-\pi/2)^{2}+\Gamma^{r}_{\;\phi\phi}(\phi-\beta t)^{2}\right]+ \cdots,\label{S16}
\end{eqnarray}
where the $\Gamma^{\alpha}_{\,\beta\gamma}$ are the Schwarzschild connection coefficients evaluated at $r=r_{0}$ and $\theta =\pi/2$. It is straightforward to compute higher order terms, but our purpose here is merely illustration of method.\\

\noindent Following Remark 4, we see from Eq. \eqref{S13} that
\begin{equation}
\left|x^{0}-\alpha t\right|\approx\frac{l}{r_{0}}r\left|\phi-\beta t\right|.\label{S17}
\end{equation}
The left side of Eq. \eqref{S17} vanishes at points on the circular orbit and increases with azimuthal deviation from $\beta t$ off the orbit and also with increasing radial distance from the central mass.  To second order, a point with Schwarzschild coordinates $(t,r,\theta,\phi)=(t,r,\theta,\beta t)$ lies on the space slice consisting of all points (sufficiently close to $\sigma$) with Fermi time coordinate $x^{0}=\alpha t$.  Thus, ignoring higher order corrections, we see that the two-dimensional surface consisting of fixed time coordinate $t$ and fixed $\phi = \beta t$ is simultaneous both for Schwarzshild and Fermi observers, i.e., it lies in the intersection of the Fermi and Schwarzshild space slices at Schwarzschild time $t$.  As $r_{0} \rightarrow \infty$, $x^{0} \rightarrow t$ and the simultaneous events in Fermi coordinates become simultaneous for the Schwarzschild observer at spacelike infinity, as expected.\\

\noindent As an application of Eq. \eqref{nM8++} consider the vector field,

\begin{equation}
\label{S19}
V = \frac{\epsilon}{X} \frac{\partial}{\partial t}+\frac{l}{r_{0}^{2}} \frac{\partial}{\partial \phi}=\left(\frac{\epsilon}{X},0,0,\frac{l}{r_{0}^{2}}\right),
\end{equation}

\noindent  with $\epsilon$ and $l$ given by Eqs. \eqref{S4}.  The right side of Eq. \eqref{S19} gives the components, $V^{\alpha}$, relative to the \textit{a priori} -- in this case, Schwarzschild -- coordinates.  Since $V$ is a linear combination of the Killing vectors $\partial/\partial t$ and $\partial/\partial \phi$, with constant coefficients, it is also a Killing vector.  Moreover, on the orbit $\sigma(t)$ given by Eq. \eqref{S2}, $V$ is given by,

\begin{equation}
V=\left(\frac{dt}{d\tau}, 0,0,\frac{d\phi}{d\tau}\right).\label{S18}\\
\end{equation}

\noindent Thus, $V$ is the four-velocity of the observer following the circular geodesic $\sigma (t)$. Using Eq. \eqref{nM8++} we easily obtain the components $V^{A}$ in Fermi coordinates as,
\begin{equation}
\label{S20}
V(\tau,x^{1},x^{2},x^{3})=\left(1,\,-\sqrt{\frac{M}{r_{0}^{3}}}\;x^{3},\,0,\,\sqrt{\frac{M}{r_{0}^{3}}}\;x^{1}\right)+O(3).
\end{equation}

\noindent The observer along $\sigma(\tau)$, using Fermi normal coordinates for measurements, may define the energy of a nearby test particle following a geodesic path with four momentum $p_{A}(\tau,x^{1},x^{2},x^{3})$ to be $-V^{A}p_{A}$.\\

\noindent \textbf{Example 2}.  The de Sitter metric is given by,
\begin{equation}
ds^{2}=-dt^{2}+e^{2Ht}\left[d(y^{1})^{2}+d(y^{2})^{2}+d(y^{3})^{2}\right]\label{CM0}
\end{equation}
where $H$ is a constant.
Consider the timelike path $\sigma(t)=(t,y^{1},y^{2},y^{3})$ with fixed \textit{a priori} space coordinates $(y^{1},y^{2},y^{3})$.  In Fermi coordinates, this path is parameterized as $\sigma(\tau) = (\tau,0,0,0)$.  Note that from Eq. \eqref{CM0}, $\tau = t$. An orthonormal tetrad along $\sigma(\tau)$ is given by,
\begin{eqnarray}
e_{0}&=&(1,0,0,0),\label{CM1}\\
e_{1}&=&(0,e^{-Ht},0,0),\label{CM2}\\
e_{2}&=&(0,0,e^{-Ht},0),\label{CM3}\\
e_{3}&=&(0,0,0,e^{-Ht}).\label{CM4}
\end{eqnarray}

\noindent This tetrad is parallel transported along the observer's geodesic (Chicone and Mashhoon \cite{Chic}).  The connection coefficients are
\begin{equation}
\Gamma^{i}_{\;it}=H,\;\;\;\;\;\Gamma^{t}_{\;ii}=e^{2Ht}H,\;\;\;\;\;i=1,2,3,\; (\mbox{for}\;\; y^{1},y^{2},y^{3}).\label{CM5}
\end{equation}

\noindent The $O(4)$ contribution in Eqs. \eqref{t13'} and \eqref{t8} is given by
\begin{multline}
\frac{1}{4!}\left(\nabla_{\nu}\nabla_{\mu}\nabla_{\beta}\delta^{(\alpha )}_{\gamma}\right) e_{I}^{\nu}e_{J}^{\mu}e_{K}^{\beta}e_{L}^{\gamma}x^{I}x^{J}x^{K}x^{L}=\\
\shoveleft{\left(-\Gamma^{(\alpha )}_{\; \gamma\beta , \mu , \nu}+4\Gamma^{(\alpha )}_{\; \sigma\beta ,\nu}\Gamma^{\sigma}_{\; \gamma\mu}
+2\Gamma^{(\alpha )}_{\; \sigma\beta}\Gamma^{\sigma}_{\; \gamma\mu ,\nu}
+\Gamma^{(\alpha )}_{\; \gamma\beta ,\lambda}\Gamma^{\lambda}_{\; \nu\mu}\right.} \\ \left. {}-
4\Gamma^{(\alpha )}_{\; \sigma\beta}\Gamma^{\sigma}_{\; \lambda\gamma}\Gamma^{\lambda}_{\; \nu\mu}+2\Gamma^{(\alpha )}_{\; \sigma\lambda}\Gamma^{\sigma}_{\; \gamma\mu}\Gamma^{\lambda}_{\; \beta\nu}\right) e_{I}^{\nu}e_{J}^{\mu}e_{K}^{\beta}e_{L}^{\gamma}x^{I}x^{J}x^{K}x^{L}.\label{CM5}
\end{multline}

\noindent Using Eq. \eqref{t8} and \eqref{CM5} above, we obtain after simple calculation (a large number of terms vanish)
\begin{eqnarray}
t=y^{t}&=&\tau -\frac{1}{H}\left[\frac{(HR)^{2}}{2}+\frac{(HR)^{4}}{12}+O(6)\right],\label{CM6}\\
y^{i}&=&e^{-H\tau}\left[x^{i} + \frac{x^{i}}{3}(HR)^{2}+O(5)\right],\label{CM&}
\end{eqnarray}

\noindent where $HR=H\sqrt{(x^{1})^{2}+(x^{2})^{2}+(x^{3})^{2}}$.  Our results above agree with Eqs. (26) and (27) of ref. \cite{Chic} to fourth order.   The method used here produces just as well Fermi-Walker coordinates for arbitrary timelike paths.\\ 

\noindent \textbf{Example 3}.  The Kerr metric in Boyer-Lindquist coordinates is given by
\begin{equation}
ds^{2}=-\frac{\rho^{2}\Delta}{\Sigma}dt^{2}+\frac{\Sigma}{\rho^{2}}\sin^{2}\theta(d\phi-\omega dt)^{2}+\frac{\rho^{2}}{\Delta}dr^{2}+\rho^{2}d\theta^{2},
\label{K1}
\end{equation}
where
\begin{eqnarray}
\rho^{2}&=&r^{2}+a^{2}\cos^{2}\theta,\label{K2}\\
\Delta&=&r^{2}-2Mr+a^{2},\label{K3}\\
\Sigma&=&\left(r^{2}+a^{2}\right)^{2}-a^{2}\Delta\sin^{2}\theta,\label{K4}\\
\omega&=&\frac{2Mar}{\Sigma}.\label{K5}
\end{eqnarray}

\noindent We consider a Zero Angular Momentum Observer (ZAMO) (or a Locally Non Rotating Frame (LNRF) \cite{BPT72}) at some fixed $r$ and $\theta$.  The ZAMO's tetrad vectors are
\begin{eqnarray}
e_{0}&=&\left(\sqrt{\frac{\Sigma}{\rho^{2}\Delta}},0,0,\frac{2Mar}{\sqrt{\rho^{2}\Delta\Sigma}}\right),\label{K6}\\
e_{1}&=&\left(0,\sqrt{\frac{\Delta}{\rho^{2}}},0,0\right),\label{K7}\\
e_{2}&=&\left(0,0,\frac{1}{\sqrt{\rho^{2}}},0\right),\label{K8}\\
e_{3}&=&\left(0,0,0,\sqrt{\frac{\rho^{2}}{\Sigma\sin^{2}\theta}}\right),\label{K9}
\end{eqnarray}

\noindent where the ordering of components is given by $(t,r,\theta,\phi)$. We now go to the equatorial plane, $\theta=\pi/2$, and fix the other coordinates $t=0$, $r=r_{0}$, $\phi = 0$. Eqs. \eqref{K6}-\eqref{K9}, evaluated at that spacetime point give the initial condition, or the $\tau=0$ value, of a Fermi-Walker transported tetrad along the circular path of a zero angular momentum observer with tangent vector $e_{0}$.  Using Eq. \eqref{t8} with $y^{0}=t,\;y^{1}=r,\;y^{2}=\theta,\;y^{3}=\phi$, and the above tetrad at $\tau=0$, we obtain
\begin{eqnarray}
t&=&-\Gamma^{t}_{\;r\phi}e_{1}^{r}e_{3}^{\phi}x^{1}x^{3}+\cdots,\label{K10}\\
r&=&r_{0}+e_{1}^{r}x^{1}-\tfrac{1}{2}\Gamma^{r}_{\;rr}\left(e_{1}^{r}\right)^{2}\left(x^{1}\right)^{2}-\tfrac{1}{2}\Gamma^{r}_{\;\theta\theta}\left(e_{2}^{\theta}\right)^{2}\left(x^{2}\right)^{2}\nonumber\\
& &-\tfrac{1}{2}\Gamma^{r}_{\;\phi\phi}\left(e_{3}^{\phi}\right)^{2}\left(x^{3}\right)^{2}+\cdots,\label{K11}\\
\theta&=&\tfrac{\pi}{2}+e_{2}^{\theta}x^{2}-\Gamma^{\theta}_{\;\theta r}e_{2}^{\theta}e_{1}^{r}x^{1}x^{2}+\cdots,\label{K12}\\
\phi&=&e_{3}^{\phi}x^{3}-\Gamma^{\phi}_{\;\phi r}e_{3}^{\phi}e_{1}^{r}x^{1}x^{3}+\cdots,\label{K13}
\end{eqnarray}
where at $\theta=\pi /2,\,r=r_{0}$, and $\Delta=\Delta(r_{0})$, we have
\begin{eqnarray}
\Gamma^{t}_{\;r\phi}&=&-\frac{aM(a^{2}+3r_{0}^{2})}{r_{0}^{2}\Delta},\;\;\;\;\;\Gamma^{r}_{\;rr}=\frac{a^{2}-r_{0}M}{r_{0}\Delta},\label{K14}\\
\Gamma^{r}_{\;\theta\theta}&=&-\frac{\Delta}{r_{0}},\;\;\;\;\;\Gamma^{r}_{\;\phi\phi}=-\frac{(r_{0}^{3}-a^{2}M)\Delta}{r_{0}^{4}},\label{K15}\\
\Gamma^{\theta}_{\;\theta r}&=&\frac{1}{r_{0}},\;\;\;\;\;\Gamma^{\phi}_{\;\phi r}=\frac{r_{0}^{2}(r_{0}-2M)-a^{2}M}{r_{0}^{2}\Delta}.\label{K16}
\end{eqnarray}\\

\noindent \textbf{Example 4}.  In this final example we calculate the Fermi-Walker coordinates $x^{A}(t,r,\theta, \phi)$ to $O(2)$ for a static observer in Kerr spacetime at fixed radial coordinate $r_{0}$ and azimuthal angle $\phi_{0}$, in the equatorial plane ($\theta=\pi/2$). We assume that the observer is outside of the ergoregion so that the four velocity $u=(dt/d\tau,0,0,0)$ is timelike.   The worldline of the observer is then $\sigma(t) = (t, r_{0}, \pi/2, \phi_{0})$.  A Fermi-Walker transported tetrad along this path was computed in \cite{maluf}.  Using the notation of Example 3, this tetrad expressed as form fields along the observer's worldline may be written,
\begin{eqnarray}
e^{0}&=&\left(\sqrt{X},0,0,\frac{2Ma}{r_{0}\sqrt{X}}\right),\label{MF1}\\
e^{1}&=&\left(0,\frac{r_{0}}{\sqrt{\Delta}}\cos(\gamma-\phi_{0}),0,\sqrt{\frac{\Delta}{X}}\sin(\gamma-\phi_{0})\right),\label{MF2}\\
e^{2}&=&\left(0,-\frac{r_{0}}{\sqrt{\Delta}}\sin(\gamma-\phi_{0}),0,\sqrt{\frac{\Delta}{X}}\cos(\gamma-\phi_{0})\right),\label{MF3}\\
e^{3}&=&\left(0,0,-r_{0},0\right),\label{MF4}
\end{eqnarray}    
where $X$ is given by Eq. \eqref{S4} and
\begin{equation}
\label{MF5}
\gamma=-\frac{aM}{r_{0}^{3}\sqrt{X}}t.
\end{equation}

\noindent The corresponding tetrad vectors are given by,
\begin{eqnarray}
e_{0}&=&\left(\frac{1}{\sqrt{X}},0,0,0\right),\label{MF6}\\
e_{1}&=&\left(-\frac{2aM\sin(\gamma-\phi_{0})}{r_{0}\sqrt{X\Delta}},\frac{\sqrt{\Delta}\cos(\gamma-\phi_{0})}{r_{0}},0,\sqrt{\frac{X}{\Delta}}\sin(\gamma-\phi_{0})\right),\label{MF7}\\
e_{2}&=&\left(-\frac{2aM\cos(\gamma-\phi_{0})}{r_{0}\sqrt{X\Delta}},-\frac{\sqrt{\Delta}\sin(\gamma-\phi_{0})}{r_{0}},0,\sqrt{\frac{X}{\Delta}}\cos(\gamma-\phi_{0})\right),\label{MF8}\\
e_{3}&=&\left(0,0,-\frac{1}{r_{0}},0\right).\label{MF9}
\end{eqnarray}

\noindent Using Eqs. \eqref{n3} and \eqref{t7} we find $x^{A}(t,r,\theta, \phi) \equiv 
x^{A}(t,y^{1},y^{2},y^{3})$ as,

\begin{equation}\label{ex4}
\begin{split}
x^{A}(t,y^{1},y^{2}&, y^{3})=\sqrt{X} \delta_{0}^{A}t+e^{A}_{k}(t)(y^{k}- y^{k}_{0})\\
&- \frac{1}{2}(\Gamma^{A}_{BC}\,e^{B}_{j}(t)e^{C}_{k}(t)-e^{A}_{\alpha}(t)\Gamma_{j k}^{\alpha})(y^{j}- y^{j}_{0})(y^{k}- y^{k}_{0})\\
& +O(3),
\end{split} 
\end{equation}

\noindent where $y^{k}_{0} = y^{k}(\sigma(t))$ is the $k$th component of $(t, r_{0}, \pi/2, \phi_{0})$ and is independent of $t$ for this example (since $k=1,2,3$), and the sum over $B$ and $C$ is carried out before the sum on $j$ and $k$ in the second term.  The connection coefficients, $\Gamma_{jk}^{\alpha}$, in Schwarzschild coordinates appearing in Eq. \eqref{ex4} are given by Eqs. \eqref{K14} --\eqref{K16}. Taking into account Eq. \eqref{F4}, the nonvanishing Fermi-Walker connection coefficients $\Gamma^{C}_{AB}$ are given by 
\begin{eqnarray}
\Gamma^{0}_{01}=\Gamma^{0}_{10}=\Gamma^{1}_{00}=a^{1}&=&\frac{M\sqrt{\Delta}\cos(\gamma-\phi_{0})}{r_{0}^{2}(r_{0}-2M)},\label{MF10}\\
\Gamma^{0}_{02}=\Gamma^{0}_{20}=\Gamma^{2}_{00}=a^{2}&=&-\frac{M\sqrt{\Delta}\sin(\gamma-\phi_{0})}{r_{0}^{2}(r_{0}-2M)}.\label{MF11}
\end{eqnarray}

\noindent We note that the reverse transformation, from Fermi-Walker to Boyer-Lindquist coordinates, may be carried out to arbitrarily high order in a straightforward manner using Eq. \eqref{t13'} along with Eqs. \eqref{MF6} -- \eqref{MF9}.\\

\noindent \textbf{{\normalsize 7. Concluding remarks}}\\

\noindent Theorem 3 gives the exact transformation formula from Fermi-Walker coordinates to arbitrary coordinates in general spacetimes. In particular, Eq. \eqref{t13'} gives explicit coefficients for expansions in terms of Fermi-Walker coordinates to arbitrarily high order. The transformation in the reverse direction is given by Theorem 2, with the Jacobian given by Eq. \eqref{j2}. Explicit formulas for vector fields were developed in Sect. 5. The examples in Sect. 6 were chosen only to illustrate our formulas, which apply quite generally to timelike paths in general spacetimes.\\

\noindent Some generalizations of the results of this paper are straightforward.  Extensions to Riemannian manifolds and to Lorentzian manifolds of arbitrary dimension $n \geqslant 2$ are easily carried out.  Using the methods of this paper, extensions to submanifolds beyond timelike paths are also possible.  A version of Theorem 3 is readily available for the case of Riemann normal coordinates on a Riemannian manifold $M$ by applying Theorem 3 to the product manifold of $M$ with the real line, with a suitable product metric.\\

\noindent Rigorous error estimates for the Taylor polynomials of the coordinate transformations developed in this paper would be useful in some circumstances, as would a rigorous lower bound for the radius of a tubular neighborhood of an arbitrary timelike path on which Fermi-Walker coordinates are valid.  Different techniques are required for the solution of those problems.\\

\end{document}